\documentclass[aps,eqsecnum,nofootinbib,showpacs,twocolumn]{revtex4}
\usepackage{amsmath,amsfonts,amssymb,bm}
\usepackage[dvips]{graphicx}
\usepackage{color}
\def\be{\begin{eqnarray}}
\def\ee{\end{eqnarray}}
\def\bec{\begin{center}}
\def\eec{\end{center}}

\def\p{\partial}
\bmdefine{\bmj}{\bm{j}}
\bmdefine{\bmk}{\bm{k}}
\bmdefine{\bmx}{\bm{x}}
\bmdefine{\bmA}{\bm{A}}
\bmdefine{\bmD}{\bm{D}}
\bmdefine{\bmF}{\bm{F}}
\newcommand{\hy}{\Hat{y}}

\newcommand{\hS}{\Hat{S}}

\newcommand{\calA}{\mathcal{A}}

\newcommand{\calC}{\mathcal{C}}

\newcommand{\calH}{\mathcal{H}}

\newcommand{\calJ}{\mathcal{J}}
\newcommand{\calL}{\mathcal{L}}

\newcommand{\calO}{\mathcal{O}}
\newcommand{\odiff}[2]{ \frac{d #1}{d #2} }
\newcommand{\odiffII}[2]{ \frac{d^2 #1}{d #2^2} }
\newcommand{\pdiff}[2]{ \frac{\partial #1}{\partial #2} }
\newcommand{\pdiffII}[2]{ \frac{\partial^2 #1}{\partial #2^2} }

\newcommand{\Exp}[1]{\left\langle~#1~\right\rangle}

\newcommand{\mfL}{\mathfrak{L}}
\newcommand{\mfj}{\mathfrak{j}}

\newcommand{\azero}{a^{(0)}}
\newcommand{\aI}{a^{(1)}}

\newcommand{\tAdS}{\text{AdS}}
\begin{document}
\title{Vortex flow for a holographic superconductor}
\author{Kengo Maeda}
\email{maeda302@sic.shibaura-it.ac.jp}
\affiliation{Faculty of Engineering,
Shibaura Institute of Technology, Saitama, 330-8570, Japan}

\author{Takashi Okamura}
\email{tokamura@kwansei.ac.jp}
\affiliation{Department of Physics, Kwansei Gakuin University,
Sanda, 669-1337, Japan}

\date{\today}
\begin{abstract}
We investigate energy dissipation associated
with the motion of the scalar condensate
in a holographic superconductor model constructed
from the charged scalar field coupled to the Maxwell field.
Upon application of constant magnetic and electric fields,
we analytically construct the vortex flow solution,
and find the vortex flow resistance near the second-order phase transition
where the scalar condensate begins.
The characteristic feature of the non-equilibrium state agrees with
the one predicted by the time-dependent Ginzburg-Landau~(TDGL) theory.
We evaluate the kinetic coefficient in the TDGL equation
along the line of the second-order phase transition.
At zero magnetic field, the other coefficients in the TDGL equation
are also evaluated just below the critical temperature.
\end{abstract}


\maketitle

\section{Introduction}\label{sec:intro}
Much attention has been given to the application of 
the AdS/CFT~(anti-de Sitter/conformal field theory)
duality \cite{Maldacena:1997re} to condensed matter physics
after discovery of holographic superconductor models
\cite{Gubser:2008px,Hartnoll:2008vx}.
Since the AdS/CFT duality is a valuable tool 
for investigating strongly coupled gauge theories, 
the application might offer new insight into
the investigation of strongly interacting condensed matter systems
where perturbative methods are no longer available.

The holographic superconductor model
constructed by charged scalar condensate \cite{Hartnoll:2008vx}
is classified into type II superconductors,
as it possesses vortex solutions
\cite{Albash:2009ix,Albash:2009iq,Montull:2009fe,Maeda:2009vf}
in a background magnetic field.
Furthermore, it has been shown that
a triangular vortex lattice solution
is the most favorable solution thermodynamically just below
the second order phase transition
at long wavelengths \cite{Maeda:2009vf}.
As already seen
in Refs. \cite{Maeda:2008ir,Hartnoll:2008kx,Maeda:2009wv},
these equilibrium states are described
by the Ginzburg-Landau~(GL) theory.
This suggests that non-equilibrium states of
the holographic superconductor in the background magnetic field
are also described by the time dependent Ginzburg-Landau~(TDGL) theory.
Indeed, it has been observed that the dynamics in the absence 
of magnetic field is described by the TDGL theory
\cite{Maeda:2009wv,AmadoKaminskiLandsteiner2009,mkf2009}.

Motivated by this, we investigate the non-equilibrium
steady state of the vortex lattice solution \cite{Maeda:2009vf}
in the presence of a small constant electric field $E$.
According to the TDGL theory, the vortex flows at a constant velocity
in a direction perpendicular to both the magnetic
and electric fields so that the Lorentz force on the vortex
is balanced by the background electric force.
The energy dissipation associated with the vortex motion occurs
in the core of the vortex~(vortex flow resistance),
as the superconducting state disappears there.

In the TDGL equation,
the evaluation of the kinetic coefficient $\Gamma$
(for example, see Eq.(\ref{TDGL-eq})) 
is important
in observing the dissipation process
or the spectrum of quasi-particles around the core of the vortex.
While it is generically difficult to derive the coefficient
from the microscopic point of view
in strongly interacting condensed matter systems,
it can be evaluated in the holographic model.
So, it is interesting to explore the dissipation mechanism
associated with the vortex motion
in the framework of the AdS/CFT duality.

In this article, we perturbatively construct the vortex flow solution 
as a series expansion of the small electric field $E$
and derive the R-current just below the critical temperature
where the scalar condensate begins.
We find that vortices flow at a constant velocity and that
the Ohmic dissipation occurs associated with the vortex motion,
as predicted by the TDGL theory. The kinetic coefficient $\Gamma$
is evaluated along the line of the second order phase transition
in comparison with the TDGL theory.
In the absence of magnetic field,
we also derive the other coefficients in the TDGL equation
from the value of the the scalar condensate, the correlation length, 
and the London equation~\cite{Maeda:2008ir}.

The plan of our paper is as follows:
In Sec.~II, we expand equation of motion for the scalar field
as a series in $E$. 
In Sec. III, we perturbatively construct the vortex flow solution 
by Green function method. 
In Sec. IV, we derive the net R-current by solving Maxwell equation
and evaluate the kinetic coefficient. 
We briefly review the TDGL theory in Appendix B, and
the other coefficients in the TDGL equation are derived in Appendix C.
Conclusions and discussion are devoted to Sec. V.
\section{Basic equations in Eddington-Finkelstein form}
\label{sec:solution}
We consider the (2+1)-dimensional holographic superconductor model 
described by a dual gravitational theory in four dimensions ($\tAdS_4$)
coupled to a charged complex scalar field $\Psi$ and a Maxwell field
$A_\mu$~\cite{Gubser:2008px,Hartnoll:2008vx}.
For simplicity, we take a probe limit
where the backreaction of the matter field onto the geometry
can be ignored \cite{Hartnoll:2008vx}.

The background metric is given by $\tAdS_4$-Schwarzschild black hole
with metric
\begin{subequations}
\begin{align}
  & ds^2 = \frac{L^2 \alpha^2}{u^2} ( - h(u) dt^2 + dx^2 + dy^2 )
  + \frac{L^2 du^2}{u^2 h(u)},
\label{eq:line_element} \\
  & h(u) = 1 - u^3,
  \hspace{1.0truecm}
  \alpha(T) = 4 \pi T/3~,
\label{eq:def-h_alpha}
\end{align}
\label{Schwarzschild-metric}%
\end{subequations}
where $L$ and $T$ are the AdS radius and the Hawking temperature,
respectively. 
We take the coordinate $u$ such that the AdS boundary is located
at $u=0$ and the horizon is set to be $u=1$. 

Under the probe limit, the action of the matter system
$S = (L^2/2 \kappa_4^2 e^2) \hS$ is written by
\begin{align}
  & \hS
  = \int d^4x~\sqrt{-g} \left(
  - \frac{F^2}{4} - |D \Psi |^2 - m^2 | \Psi |^2 \right),
\label{action1}
\end{align}
where $m$ and $e$ are the mass and charge of the scalar field $\Psi$, 
respectively, and
\begin{align}
  & D_\mu := \nabla_\mu - i A_\mu~,
& & F_{\mu\nu} := \p_\mu A_\nu - \p_\nu A_\mu~.
\label{eq:def-cov_deri_D}
\end{align}
Hereafter, we consider the action~(\ref{action1}) in the 
simple case $m^2L^2=-2$. The equations of motion are given by 
\begin{subequations}
\begin{align}
  & D^2 \Psi +\frac{2}{L^2} \Psi = 0~,
\label{eq:EOM-Psi} \\
  & \nabla_\nu F_\mu{}^\nu = j_\mu
  := i [ (D_\mu\Psi)^\dagger \Psi - \Psi^\dagger (D_\mu\Psi) ]~.
\label{eq:EOM-A}
\end{align}
\label{eq-motion}
\end{subequations}

For a gauge choice, we choose a gauge $A_u=0$
in the metric~(\ref{Schwarzschild-metric}).
The asymptotic behavior of $\Psi$ and $A_\mu~(\mu=t,x,y)$
near the AdS boundary are
\begin{subequations}
\label{eq:bc-bdy}
\begin{align}
   \Psi
  &\simeq
  c_1(t, x, y)\, u + c_2(t, x, y)\, u^{2}~,
\label{eq:bc_Psi-bdy} \\
   A_\mu
  &\simeq
  \calA_\mu(t, x, y) + \calJ_\mu (t, x, y)\, u~. 
\label{eq:bc_A-bdy}
\end{align}
\end{subequations}
According to the AdS/CFT dictionary,
the expectation values of the dual scalar operator ${\cal O}_2$%
\footnote{Our definition of $\Exp{\calO_2}$ is the same as
Ref.~\cite{Herzog:2008he} in units where $L^2/2 \kappa_4^2 = 1$. }
with conformal dimension two and the R-current $J^\mu~(\mu=t, x, y)$ are 
represented by the coefficient $c_2$ and $\calJ_\mu$ as
\begin{subequations}
\label{eq:expectation}
\begin{align}
  & \Exp{\calO_2}
  = \left. \frac{\sqrt{2}\, e}{L\, \alpha}\,
  \frac{\delta S}{\delta c_1^\dagger}\, \right\vert_{u=0}
  = \frac{\sqrt{2}\, L^3\, \alpha^2}{2\, \kappa_4^2\, e}\, c_2~,
\label{eq:expectation-o} \\
  & \Exp{J^\mu}
  = \left. \frac{\delta S}{\delta \calA_\mu}\, \right\vert_{u=0}
  = \frac{L^2\, \alpha}{2\, \kappa_4^2\, e^2}\,
  \eta^{\mu\nu}\, \calJ_\nu~,
\label{eq:expectation-j}
\end{align}
\end{subequations}
respectively. We consider the condensation of the scalar operator 
${\cal O}_2$, and impose an asymptotic boundary condition $c_1=0$ to 
eliminate the source term in the dual theory. 

Since we are interested in the superconducting region
just below the second order phase transition,
the amplitude of the scalar field $|\Psi|$ is very small.
So, one can expand $\Psi$ and $A_\mu$ in powers of a small parameter
$\epsilon$ as 
\begin{align*}
  & \Psi = \epsilon^{1/2}\, \psi + O(\epsilon^{3/2})~,
& & A_\mu = \bmA_\mu + \epsilon\, a_\mu  + O(\epsilon^2)~. 
\end{align*}
Then, Eq.~(\ref{eq:EOM-A}) at $O(\epsilon^0)$ 
is reduced to 
\begin{align}
  & \nabla_\nu \bmF_\mu{}^\nu = 0
\label{eq:zeroth_EOM-A},  
\end{align}
where $\bmF_{\mu\nu}=\p_\mu\bmA_\nu-\p_\nu\bmA_\mu$. 
By Eqs.~(\ref{eq-motion}), the equations of motion
for $\psi$ and $a_\mu$ become 
\begin{subequations}
\label{eq:perturbed_EOM}
\begin{align}
  & \bmD^2 \psi +\frac{2}{L^2} \psi = 0~,
\label{eq:perturbed_EOM-Psi} \\
  & 
  \nabla_\nu f_\mu{}^\nu
  = i\, \left[~(\bmD_\mu\psi)^\dagger \psi
  - \psi^\dagger (\bmD_\mu\psi)~\right]~,
\label{eq:perturbed_EOM-A}
\end{align}
\end{subequations}
where $\bmD_\mu$ and $f_{\mu\nu}$ are defined by
$\bmD_\mu := \partial_\mu - i\, \bmA_\mu$ and
$f_{\mu\nu} := 2\, \partial_{[\mu} a_{\nu]}$, respectively.

We consider a zeroth order solution of Eq.~(\ref{eq:zeroth_EOM-A}) 
generating a constant electric field $E$
and the (upper) critical magnetic field $B_{c2}$
at the second order phase transition \cite{Maeda:2009vf}.
This is given by the following form:
\begin{align}
  & \bmA_t = \mu (1 - u)~,
& & \bmA_x
  = - E\, (t-u_*) - B_{c2}\, y~,
\label{gauge-choice}
\end{align}
where $u_* := \int^u du/\alpha h(u)$
and $\mu$ is the chemical potential.
The boundary conditions at the horizon $u=1$ are determined
by the regularity condition, i.e.,
$\bmA_t(x,y,u=1)=0$ and $|\bmF^2(x,y,u=1)|<\infty$.
Our strategy is to solve Eqs.~(\ref{eq:perturbed_EOM}) perturbatively
for small $E$ under the external gauge field~(\ref{gauge-choice}).
In the $E = 0$ case, Eq.~(\ref{eq:perturbed_EOM-Psi}) is solved
as the Landau problem \cite{Albash:2008iv},
and Eq.~(\ref{eq:perturbed_EOM-A}) is also formally
solved \cite{Maeda:2009vf}.

Let us expand Eqs.~(\ref{eq:perturbed_EOM}) in powers of $E$ as
\begin{subequations}
\label{eq:perturbed_E}
\begin{align}
  & \psi
  = \psi_0 + E\, \psi_1 + \cdots~,
\label{eq:perturbed_E-Psi} \\
  & a_\mu
  = \azero_\mu + E\,\aI_\mu + \cdots~. 
\label{eq:perturbed_E-A}
\end{align}
\end{subequations}
As shown later,
it is convenient to adopt an advanced null coordinate
$v := t-u_*$
and a coordinate $\hat{y}$ defined by
\begin{align}
\label{new-y-coordinate}
  & \hat{y}
  :=y+ \frac{E}{B_{c2}}\, v~.
\end{align}
%
Under the coordinate transformation $(t, u, x, y) \mapsto
(v, u, x, \hy)$, 
the metric (\ref{Schwarzschild-metric})
and the gauge field (\ref{gauge-choice}) are transformed as
\begin{subequations}
\begin{align}
  & ds^2
  = \frac{L^2\alpha^2}{u^2} \bigg( - h(u)\, dv^2
  - \frac{2}{\alpha}\, dv\, du + dx^2 + d\hat{y}^2
\nonumber \\
  &\hspace*{2.0truecm}
  - \frac{2 E}{B_{c2}}\, d\hat{y}\, dv
  \bigg) + O(E^2)~,
\label{eq:Eddington-coordinate} \\
  & \bmA_v=\mu(1-u), \quad
  \bmA_{u}
  = \frac{\bmA_v}{\alpha\, h(u)}, \quad 
  \bmA_x=-B_{c2}\,\hat{y}.
\label{eq:new-gauge-field}
\end{align}  
\end{subequations}
Thus, the external electric field $E$ appears only
via the metric in the Eddington-Finkelstein form
(\ref{eq:Eddington-coordinate}).
%

If the holographic superconductor obeys
conventional type II superconductors,
the vortex flows at a constant velocity $-E/B_{c2}$
along the $y$-direction by the Lorentz force.
This implies that there is a ``static" vortex solution
in the metric (\ref{eq:Eddington-coordinate}). 
So, we take an ansatz for the scalar field $\psi$%
\footnote{In Ref.~\cite{Maeda:2009vf},
the expansion along the $x$-direction is given by Fourier series
rather than the Fourier transform.
As shown in the Appendix A, however,
we can formally write the vortex lattice solution
in the framework of the Fourier transform.}:
%
\begin{subequations}
\label{ansatz-scalar}
\begin{align}
  & \psi(x,\hat{y},u)
  = \int^\infty_{-\infty} \frac{dp}{\sqrt{2 \pi}}~
  C(p)\, e^{i p x}\, \xi(\hat{y},u;p),
\nonumber \\
&\xi(\hat{y},u;p)
  = \xi_0(\hat{y},u;p)
  + E\, \xi_1(\hat{y},u;p) + \cdots~.
\label{ansatz-scalar1}
\end{align}
\end{subequations}
%

Defining the differential operators $\mfL_p$ and $\calL$ as 
\begin{subequations}
\begin{align}
  & \mfL_p
  := \calL
  + \frac{1}{\alpha^2}\left[\frac{\p^2}{\p \hat{y}^2}
  - (p+B_{c2}\,\hat{y})^2 \right],
\label{eq:def-mfL} \\
  & \calL
  := u^2 \pdiff{}{u} \frac{h(u)}{u^2} \pdiff{}{u}
  + \frac{\mu^2(1-u)^2}{\alpha^2\, h(u)}
  +\frac{2}{u^2}~,
\label{eq:def-calL}
\end{align}
\end{subequations}
we obtain the equations of motion for $\xi_0$ and $\xi_1$
from Eq.~(\ref{eq:perturbed_EOM-Psi}) as
\begin{subequations}
\begin{align}
  & \mfL_p\, \xi_0=0~,
  \hspace{2.0truecm} \mfL_p\, \xi_1= \mfj~,
\label{psi0_psi1} \\
  & \mfj
  := 
  \frac{2}{\alpha B_{c2}}\, \pdiff{}{\Hat{y}}
  \left( u\, \pdiff{}{u}\, \frac{1}{u}
    - \frac{i \mu (1-u)}{\alpha h(u)} \right) \xi_0~.
\label{def_j1}
\end{align}
\label{eq:xi0xi1}
\end{subequations}
%
The boundary conditions of $\Psi$ are represented by 
\begin{align}
\label{boundary-con}
  & \Psi(x, \hy, u)
  = \begin{cases}~c_2(x, \hy)\, u^2  & (u\to 0)
  \\ ~\text{regular} & (u\to 1)
\end{cases}~.
\end{align}   
\section{The construction of the vortex flow solution}
In this section, we construct the solutions $\xi_0$ and $\xi_1$ 
of Eqs.~(\ref{psi0_psi1}). 
Following the ansatz in Ref.~\cite{Maeda:2009vf}, we separate 
the variable $\xi_0$ as $\xi_0= \rho_n(u)D_n(Y_p)$, where $Y_p$ 
is defined by  
\begin{align}
Y_p := \sqrt{2B_{c2}}\left(\hat{y}+\frac{p}{B_{c2}}\right). 
\end{align} 
The equations for $\rho_n(u)$ and $D_n(Y_p)$ are derived from 
Eq.~(\ref{psi0_psi1}) as
\begin{subequations}
\begin{align}
  & \calL\, \rho_n(u)
  = \frac{B_{c2}\lambda_n}{\alpha^2}\, \rho_n(u)~,
\label{eq-rho0} \\
  & \left( \pdiffII{}{Y_p} - \frac{Y_p^2}{4} \right) D_n(Y_p)
  = - \frac{\lambda_n}{2}\, D_n(Y_p), 
\label{eq:D_n}
\end{align}
\end{subequations}
where $\lambda_n$ is a separation constant.
The solution of the equation~(\ref{eq:D_n}) satisfying
the boundary condition
$\lim_{|Y|\to \infty}|D_n(Y)|<\infty$
is given by 
\begin{subequations}
\begin{align}
  & \lambda_n = 2 n + 1  \hspace{1.0truecm} (n = 0, 1, 2, \cdots)~, 
\label{sol:lambda_n} \\
  & D_n(Y)
  := \left(\frac{1}{ \sqrt{2 \pi}\, 2^n n!}\right)^{1/2}\,
  H_n\left( \frac{Y}{\sqrt{2}} \right)\, e^{-Y^2/4}~,
\label{sol:D_n}
\end{align}
\end{subequations}
where $H_n$ is the Hermite function defined by 
\begin{align}
\label{hermite}
H_n(z):=(-1)^ne^{z^2}\frac{\p^n}{\p z^n}(e^{-z^2}).  
\end{align} 

The function $D_n(Y_p)$ in Eq.~(\ref{sol:D_n})
is the $n$-th energy eigenfunction of a harmonic oscillator centered at
$\hat{y} =- p/B_{c2}$
and it exponentially decays for large $|Y_p|$.
As discussed in Ref.~\cite{Maeda:2009vf},
the upper critical value $B_{c2}$ is determined by $n=0$ and
the solution $\rho_0$
satisfying the two boundary conditions~(\ref{boundary-con})
was numerically obtained.
Therefore, we shall adopt $n=0$ solution, 
$\xi_0 = \rho_0(u)\, D_0(Y_p)$ as the leading order solution 
of Eq.~(\ref{psi0_psi1}). 

We derive the next order solution $\xi_1$ of Eq.~(\ref{psi0_psi1})
by constructing Green function.
In general, $\xi_1$ includes a component proportional to $\xi_0$.
Hereafter, we shall remove this component from $\xi_1$
because it can be absorbed into the leading order solution $\xi_0$. 

Introducing the inner product for $D_n$ as
\begin{align}
\label{con:complete-orthonormal}
  & (\xi,\,\eta)
  := \int_{-\infty}^\infty dY~\xi^\dagger(Y)\, \eta(Y)~,
\end{align}
$\{D_n\}$ forms a complete orthonormal set%
\begin{subequations}
\begin{align}
  & (D_n,\,D_m) = \delta_{nm}~,
\label{orthogonality} \\
  & \sum_{n=0}^\infty
  D_n(Y_p) D_n^\dagger(Y'_p) = \delta(Y_p - Y'_p )
  = \frac{ \delta(\hat{y}-\hat{y}')}{\sqrt{2 B_{c2}}}~.
\label{completeness}
\end{align}
\end{subequations}
As well known, $\{D_n\}$ satisfies the relation 
%
\begin{align}
\label{formula}
   (D_n,\,YD_m)
  &= \begin{cases}
    \sqrt{n+1} & ~~m=n+1
  \\
    \sqrt{n} & ~~m=n-1
  \\
    0 & ~~|m-n|\neq 1
  \end{cases}~{\color{red}.}
\end{align} 
In terms of the complete orthonormal set $\{D_n\}$, let us construct 
the Green function $G_p(u, \hy\, |\, u', \hy'\,)$ of the operator
$\mfL_p$ in the form
\begin{align}
  & \frac{ G_p(u, \hy\, |\, u', \hy'\,) }{ \sqrt{2B_{c2}} }
  = \sum_{n=1}^\infty\, g_n(u, u' )\, D_n(Y_p)\, D_n^\dagger(Y_p')~. 
\label{sol-greenG} 
\end{align}
Suppose that the two point function $g_n$~$(n = 1, 2, \cdots)$
is the solution of the equation
\begin{align}
  & \left( \calL - \frac{B_{c2} \lambda_n}{\alpha^2} \right)
  g_n(u, u'\,) = \delta(u-u')~
\label{def-green-gn}
\end{align}
satisfying the boundary condition (\ref{boundary-con}). 
Then, using the completeness of $D_n(Y)$ in Eq.~(\ref{completeness}),
we find that $G_p$ satisfies
\begin{align}
  & \mfL_p\, G_p(u, \hy\, |\, u', \hy'\,)
  = \delta( \hy - \hy'\,) \delta(u-u'\,)
\nonumber \\
  &\hspace*{1.0truecm}
  - \sqrt{2 B_{c2}}\, \delta(u - u'\,)\, D_0(Y_p)\,
    D_0^\dagger(Y_p')~.
\label{def-greenG}
\end{align}
This indicates that $G_p$ is the Green function in the solution space
orthogonal to the component $\xi_0\propto D_0$. 

%
%

In terms of the Green function $G_p$, the formal solution of $\xi_1$
is represented as
\begin{align}
\label{sol-formal-psi1}
  & \xi_1(\hy, u ; p)
  =
   \int du'd\hy'~G_p(u, \hy\, |\, u', \hy'\,)\, \mfj(\hy', u' ;p )~.
\end{align}
%
Substituting Eq.~(\ref{sol-greenG})
into the solution (\ref{sol-formal-psi1})
and using the orthogonality condition (\ref{orthogonality}), 
one obtains
\begin{align}
\label{sol-psi1}
  & \xi_1(\hy, u; p)
  = 
  - \frac{2}{\alpha\, \sqrt{2 B_{c2}} }\, D_1(Y_p)
\nonumber \\
  &
  \times
  \int^{1}_0 du'~g_1(u, u'\,)
    \left( u' \odiff{}{u'}\, \frac{1}{u'}
    - \frac{i \mu (1-u')}{\alpha h(u')} \right) \rho_0(u')
\nonumber \\
  &=: 
  - \frac{D_1(Y_p)}{\alpha\, \sqrt{2 B_{c2}} }\,
  \big\{ \rho_R(u) + i\, \rho_I(u) \big\}~,
\end{align}
where $\rho_R$ and $\rho_I$ are the real and imaginary parts
of the $u'$-integral, respectively. 
\section{Energy dissipation associated with the vortex flow}
\label{sec:DC conductivity}
In this section, we investigate energy dissipation caused by the vortex flow 
solution constructed in the previous section. As shown below, the R-current 
associated with the vortex flow agrees with the one predicted by the TDGL theory~(see, Appendix B). 
So, we can evaluate the kinetic coefficient $\Gamma$ in the TDGL equation~(\ref{TDGL-eq}) 
from the R-current. The other coefficients are also evaluated from our earlier 
results~\cite{Maeda:2008ir} in Appendix C. We begin by calculating the R-current 
induced by the vortex flow.  

\subsection{R-current}
$\Exp{J^x}$ in Eq.~(\ref{eq:expectation-j}) can be expanded
as a series in $\epsilon$ near the second order phase transition as
\begin{align}
  & \Exp{J^x}
  = \frac{L^2E}{2\kappa_4^2 e^2} + \delta \Exp{J^x}~,
\end{align}
where $\delta \Exp{J^x} = O(\epsilon) = O(\Psi^2)$.
The first term is caused by the normal fluid,
which is independent of the vortex motion.
Hereafter, we will calculate the subleading term $\delta \Exp{J^x}$,
as it is induced by the vortex motion.

For simplicity,
we shall focus attention on calculating the net~(total) current
$\overline{\delta \Exp{J^x}}$.
We define the net value $\bar{A}$ of a quantity $A$
in the original coordinate $(x,y)$ as
\begin{align}
\label{def-average}
\bar{A} := \iint dx dy~A.
\end{align}
By Eqs.~(\ref{eq:bc_A-bdy}) and (\ref{eq:expectation-j}), 
$\overline{\delta \Exp{J^x}}$ becomes 
%
\begin{align}
\label{average}
   \overline{\delta \Exp{J^x}}(t)
  &= \frac{L^2 \alpha \epsilon}{ 2 \kappa_4^2{e^2}}\,
  \iint dx dy~\partial_u a_x
  \big\vert_{u=0}
\nonumber \\
  &= \frac{L^2 \alpha \epsilon}{ 2 \kappa_4^2{e^2}}\,
  \partial_u \bar{a}_x
  \big\vert_{u=0}~.
 \end{align}

We first derive $\bar{a}_x$ by solving Eq.~(\ref{eq:perturbed_EOM-A}).
As in Eq.~(\ref{eq:perturbed_E}),
the net bulk current $\bar{j}_x$ can be expanded as
\begin{align}
  & \bar{j}_x
  = \bar{j}^{(0)}_x
  + E\, \bar{j}^{(1)}_x + \cdots~.
\label{bulk-current-expansion}
\end{align}
The leading term $\bar{j}^{(0)}_x$ is clearly zero
because the current circulates~(see, Ref.~\cite{Maeda:2009vf}) as
\begin{align}
 & j_x^{(0)}
  = - \epsilon\,
  \partial_y |\psi_0|^2.  
\label{current-component}
\end{align} 
This implies that the leading term $\bar{a}^{(0)}_x$ in
Eq.~(\ref{eq:perturbed_E-A}) is zero, and that $\bar{a}_x$ 
is written by $\bar{a}_x = E\, \bar{a}^{(1)}_x + \cdots$. 

By Eqs.~(\ref{orthogonality}), (\ref{formula}), and (\ref{sol-psi1}),
the subleading term $\bar{j}^{(1)}_x$ is calculated as
\begin{subequations}
\label{averaged-source}
\begin{align}
   \bar{j}^{(1)}_x
  &= - 2\epsilon \overline{\Im\left[
  (D_x\psi_0)^\dagger \psi_1 + (D_x\psi_1)^\dagger \psi_0 \right] }
\nonumber \\
  &= - \frac{2 \epsilon\,\calC}{ \alpha \sqrt{2 B_{c2}} }\,
  \rho_0(u)\, \rho_R(u)~,
\label{source-x} \\
   \calC
  &:= \int dp~|C(p) |^2~,  
\end{align}
\end{subequations}
which is independent of the coordinate $v$. In the spirit of finding
the solution of Eq.~(\ref{eq:perturbed_EOM-A}), we assume that 
$\bar{a}^{(1)}_\mu$ is also $v$-independent. 
Then, averaging the $x$-component of Eq.~(\ref{eq:perturbed_EOM-A})
and using Eq.~(\ref{averaged-source}), we find
\begin{align}
   \odiff{}{u}\, \left( h\, \odiff{\bar{a}^{(1)}_x(u)}{u} \right)
  &= \frac{2\, L^2\,\calC }{\alpha \sqrt{2 B_{c2}}  u^2}\,
  \rho_0(u)\, \rho_R(u). 
\end{align}
Here, we used the fact that the spatially averaged value
in Eq.~(\ref{def-average}) is  zero for the derivative terms of $a_\mu$
with respect to the spatial coordinates, $x$, $y$,
i.~e.~, $\overline{\partial_x a_\mu}=\overline{\partial_y a_\mu}=0$.
The regularity of $\bar{a}^{(1)}_x$ at the horizon
determines the solution as
\begin{align}
\label{sol-current-x}
  \odiff{}{u}\, \bar{a}_x^{(1)}(u)
  = - \frac{2\, L^2\, \calC}{\alpha \sqrt{2 B_{c2}} h(u)}\, 
\int^1_u du'~\frac{\rho_0(u')\, \rho_R(u')}{u'^2}~.
\end{align}
Substituting Eq.~(\ref{sol-current-x}) into Eq.~(\ref{average}), 
the net R-current becomes
\begin{align}
  & \overline{\delta\Exp{{J}^x}}
  = -
  \frac{\epsilon\,E\,L^4\,\calC}{\kappa_4^2 {e^2}\, \sqrt{2 B_{c2}} }\,
  \langle\rho_0,\,\rho_R\rangle, 
\label{eq:bar-J_x-1st-pre}
\end{align}
where $\langle \rho_0,\,\rho_R \rangle$ is the inner product defined by
\begin{align}
  & \langle \phi,\,\psi \rangle
  := \int^1_0 du~\frac{\phi^*(u)\, \psi(u)}{u^2}. 
\label{eq:def-IP-AdS_rad}
\end{align}

Next, we express the net R-current in Eq.~(\ref{eq:bar-J_x-1st-pre})
in terms of the expectation value of the scalar operator
$\Exp{{\cal O}_2}$.
Under the boundary condition (\ref{boundary-con}), the operator $\calL$
is clearly Hermitian for the inner product (\ref{eq:def-IP-AdS_rad}).
So, using Eqs.~(\ref{eq-rho0}), (\ref{def-green-gn}),
and (\ref{sol-psi1}), we obtain the following equality:
\begin{align}
  & \langle \rho_0,\, \calL\, \rho_R \rangle
  = \frac{B_{c2}\lambda_1}{\alpha^2}\,
  \langle \rho_0,\, \rho_R \rangle
  + 2 \left\langle \rho_0,\, u\, \frac{d}{du}\, \frac{\rho_0}{u}
      \right\rangle
\nonumber \\
  =&\, \langle \calL\, \rho_0,\,\rho_R \rangle
  = \frac{B_{c2}\lambda_0}{\alpha^2}\,
  \langle \rho_0,\,\rho_R \rangle~. 
\end{align}
The boundary condition (\ref{boundary-con}) simplifies the equality as
\begin{align}
   \langle \rho_0,\,\rho_R \rangle
  = - \frac{\alpha^2}{2B_{c2}}\, \rho_0^2(u=1)~.
\label{eq:uIP-rho_zero-rho_R}
\end{align}
Substituting Eq.~(\ref{eq:uIP-rho_zero-rho_R}) into
Eq.~(\ref{eq:bar-J_x-1st-pre}),
$\overline{\delta\Exp{J^x}}$ is expressed by the expectation
value of the dual scalar operator $\Exp{{\cal O}_2}$:
\begin{align}
   \overline{\delta\Exp{J^x}}
  &= \frac{\epsilon L^4 \alpha^2}{2 \kappa_4^2{e^2}}\,
  \frac{E}{B_{c2}}\,
  \frac{\calC\, \rho_0^2(1)}{ \sqrt{2 B_{c2}} } 
  = \frac{L^4 \alpha^2}{2 \kappa_4^2{e^2}}\,
  \frac{E}{B_{c2}}\,
  \left. \overline{ |\Psi_0 |^2 }\, \right\vert_{u=1}
\nonumber \\
  &=\frac{\kappa_4^2\beta^2}{L^2\alpha^2}\frac{E}{B_{c2}}\,
  \overline{
  \left\vert \Exp{{\cal O}_2} \right\vert^2}~.
\label{eq:bar-J_x-1st}
\end{align}
Here, $\Psi_0 := \epsilon^{1/2}\, \psi_0$
and the coefficient $\beta$ is defined by
\begin{align}
  & \beta^2
  := \frac{\overline{ |\Psi_0 |^2 }\, \Big\vert_{u=1} }
  { \overline{| c_2 |^2 } }
  = 
  \frac{ |\rho_0(u=1) |^2 }
  {\lim_{u \to 0} | \rho_0(u)/u^2 |^2 }~.
\end{align}

Eq.~(\ref{eq:bar-J_x-1st}) shows that a finite DC-current is induced by 
the motion of the scalar field in a parallel direction with the applied 
electric field $E$. Thus, the vortex flow resistance appears by the vortex 
motion in the holographic superconductor model. In the bulk side, the energy 
dissipation (Ohmic dissipation) associated with the resistance is represented 
by the energy absorption of the scalar field by the black hole. As shown in 
Eq.~(\ref{eq:conserve_eqn-stationary-concrete-perturbed}), 
the energy flows into the bulk from the boundary via the external electric field. 
It is transformed into the energy of the scalar field in the bulk. Since the scalar field 
falls into the black hole horizon, as it moves in the $y$-direction, the energy is 
absorbed into the black hole.

\subsection{The kinetic coefficient $\Gamma$}
The form of the expectation value (\ref{eq:bar-J_x-1st}) agrees
with the averaged value of the current in TDGL theory (\ref{App:current}).
Then, the kinetic coefficient $\Gamma$ is given by
\begin{align}
  & \frac{\Gamma(T)}{e_*}
  = 
  \left. \frac{L^2 \mu^2}{\kappa_4^2}\,
  \frac{(\alpha/\mu)^2}{\beta^2}\,
  \right\vert_{T, B=B_{c2}(T)}
  =: \frac{L^2 \mu^2}{\kappa_4^2}\, Z~. 
\label{Gamma/e}
\end{align}

We can easily show that $e_*=1$ in the following argument.  
As seen in Eq.~(\ref{eq:def-cov_deri_D}), 
the gauge coupling between $A_\mu$ and $\Psi$ is given in the form,   
$(\partial_\mu - i\, A_\mu) \Psi$. Under the gauge $A_u = 0$, 
there is still a residual gauge transformation~\cite{Maeda:2010br}:  
\begin{subequations}
\label{eq:bulk-residual_gauge_tr}
\begin{align}
   A_\mu(t, x, y, u)
  &\to A_\mu(t, x, y, u) + \partial_\mu \Lambda(t, x, y)~,
\\
   \Psi(t, x, y, u)
  &\to e^{i \Lambda(t, x, y)}\, \Psi(t, x, y, u). 
\end{align}
\end{subequations}
Then, Eq.~(\ref{eq:bulk-residual_gauge_tr}) acts on the source
$\calA_\mu$ of the R-current and on the condensate $\Exp{\calO_2}$
dual to $\Psi$ as
\begin{subequations}
\label{eq:bdy-gauge_tr}
\begin{align}
   \calA_\mu(t, x, y)
  &\to \calA_\mu(t, x, y) + \partial_\mu \Lambda(t, x, y)~,
\\
   \Exp{\calO_2(t, x, y)}
  &\to e^{i \Lambda(t, x, y)}\, \Exp{\calO_2(t, x, y)}~.
\end{align}
\end{subequations}
This is a \lq\lq background local U(1)\rq\rq~transformation
of the dual field theory, indicating that the gauge coupling 
constant $e_\ast$ of $\Exp{\calO_2}$ is unity,
{\it i.e.}, $e_\ast = 1$.  

By solving numerically Eq.~(\ref{eq-rho0}) with $n = 0$, 
we obtain $Z := (\alpha/\mu \beta)^2
= \kappa_4^2\,\Gamma/(L^2 \mu^2)$,  
which is a function of 
$\alpha/\mu \propto T/T_c$ only. 
In Fig.~1, we present $\Gamma(T)$ as a function of $T$. It increases
as $T$ decreases from the critical temperature $T_c$ at zero magnetic field, $B_{c2}=0$.
In the $B_{c2}\to 0$ limit~($T\to T_c$), 
both $\alpha/\mu$ and $Z$ approach critical values
$\alpha_c/\mu\sim 0.25$ and  $Z \sim 0.54$, respectively,
which are independent of the critical temperature $T_c$.
Thus, we finally obtain $\Gamma$ as   
\begin{align}
  & \Gamma(T_c)
  \sim \left( \frac{L^2}{2 \kappa_4^2} \right)
  303.2 \times T_c^2~
\label{num-Gamma/e}
\end{align}
in the $T\to T_c$ limit. 

\begin{figure}
\label{fig1}
 \begin{center}
  \includegraphics[width=8.5truecm,clip]{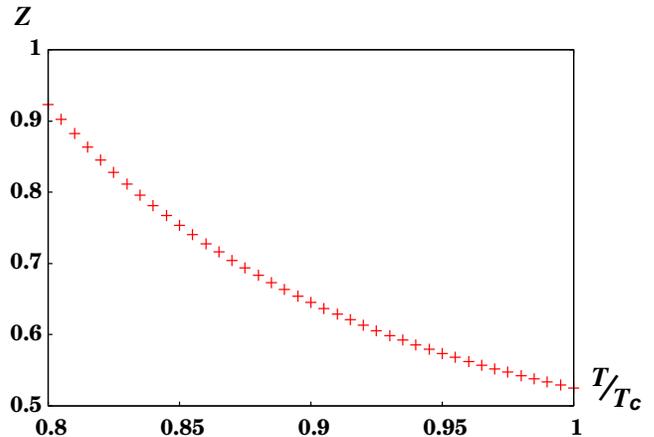}
  \caption{ 
  $T$-dependence of $Z := \kappa_4^2\,\Gamma/(L^2 \mu^2)$
  with $\mu$ fixed. Here $T_c$ is the critical temperature
  for $B_{c2}=0$.
  } 
 \end{center}
\end{figure}

\section{Conclusions and discussion}\label{sec:discussion}
We have investigated the vortex motion of a holographic superconductor
constructed by a gravitational model of complex scalar field
coupled to the $U(1)$ gauge field.
We found that the vortex flows in a direction orthogonal
to both the electric field $E$ and the magnetic field $B$
at a constant velocity $v=E/B$.
This is explained by the force balance between the Lorentz force 
and the electric force observed in the conventional 
type II superconductors \cite{parks}.

We observed Ohmic dissipation associated with the vortex motion. 
This might be explained by the speculation that
the superconducting state is violated
at each core of the vortex lattice.
In other words, the normal state at each core
causes the energy dissipation by the constant motion.
Since the DC-conductivity we calculated in Sec.~IV
is the spatially averaged value, we cannot say exactly
where the dissipation occurs in the vortex motion.
Indeed, the dissipation is independent of the 
coefficient $C(p)$ in Eq.~(\ref{ansatz-scalar}).
It is interesting to investigate further
the location of the dissipation in our model
by calculating the DC-conductivity at each point.

As shown in Sec.~IV,
the DC-current agrees with the current in the TDGL theory.
The kinetic coefficient $\Gamma$ 
in the TDGL equation was obtained along the line $B=B_{c2}(T)$
in $(B,T)$ phase diagram as shown in Fig.~2. We also obtained
the other coefficients in the TDGL equation
just below the critical temperature $T_c$ at zero magnetic field.
It is worth comparing these coefficients obtained in this article 
with the ones obtained from other phenomena as a consistency check. 

In general, there is a possibility that $\Gamma$ depends on $T$
and independently on $B$, i.~e., $\Gamma=\Gamma(T,\,B)$.
To investigate the possibility, we need to evaluate $\Gamma$
along another line away from the $B=B_{c2}(T)$ line.
It would be interesting to clarify the dependency in the phase diagram
where the TDGL theory is available, and to compare it with experiments.
Then, we might be able to find a sign of a strongly correlated
condensed matter system in the holographic superconductor model.

\begin{figure}
\label{fig2}
 \begin{center}
  \includegraphics[width=8.5truecm,clip]{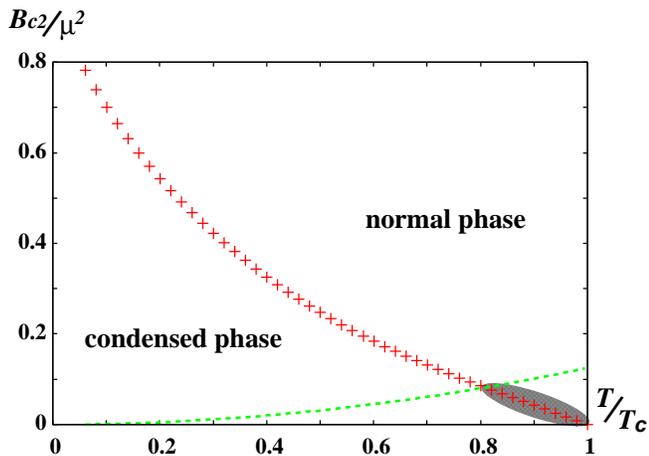}
  \caption{(color online)
  Phase diagram of the holographic superconductor.
  $B_{c2}(T)$ is marked with \lq\lq $+$\rq\rq (red).
  Since the local approximation is valid
  below the dashed line (green), as shown in Ref.~\cite{Maeda:2009vf},
  the Ginzburg-Landau theory is applicable in the shaded region.
  }
 \end{center}
\end{figure}

\begin{acknowledgments} 
We would like to thank S. Tsuchiya and M.~Natsuume
for useful discussions.
This research was supported in part by the Grant-in-Aid for Scientific
Research~(20540285) from the Ministry of Education, Culture,
Sports, Science and Technology, Japan.
\end{acknowledgments}
\appendix
\section{vortex lattice solution}
In the zero limit of the external electric field $E \to 0$,
the solution should be reduced to the static vortex lattice solution
obtained in Ref. \cite{Maeda:2009vf}.
Let us take $C(p)$ in Eq.~(\ref{ansatz-scalar}) as
\be
C(p)=\sum_{l=-\infty}^\infty\delta(p-p_l)\,C_l,
\ee
where $p_l$ and $C_l$ are defined by two lattice parameters,
$a_1$ and $a_2$:
\be
\label{def:p_l}
p_l:=\frac{2\pi\sqrt{B_{c2}} l}{a_1}, 
\qquad C_l:=\exp\left(-i\frac{\pi a_2}{a_1^2}l^2  \right).  
\ee 

In the $E \to 0$ limit, 
$\Psi$ has a pseudoperiodicity 
\begin{subequations}
\begin{align}
  & \Psi(x,\,y,\,u)
  = \Psi(x+a_1\,r_0, y,\,u),
\label{pseudo-periodicity1} \\
  & \Psi\left(x+\frac{a_2\,r_0}{a_1}, y-\frac{2\pi r_0}{a_1},\,u\right)
\nonumber \\
&\hspace{0.5truecm}
  = \exp\left[ \frac{2 \pi i}{a_1}
  \left( \frac{x}{r_0} + \frac{a_2}{2 a_1} \right) \right]
  \Psi(x,\,y,\,u).
\label{pseudo-periodicity2}  
\end{align}
\label{pseudo-periodicity}
\end{subequations}
Thus, the fundamental region $V_0$ on the boundary $u=0$ is spanned
by two vectors, ${\bm b}_1=a_1r_0\p_x$ and
${\bm b}_2=a_2r_0/a_1\p_x-2\pi r_0/a_1\p_y$.
Here $r_0:=1/\sqrt{B_{c2}}$ is the typical length scale
of the fundamental region \cite{Maeda:2009vf}.
The triangular lattice solution, for example, is given by
\be
\frac{a_2}{a_1}=\frac{a_1}{2}=3^{-1/4}\sqrt{\pi}. 
\ee
\section{The vortex flow solution to TDGL equation}
The conventional superconductors near the critical temperature $T=T_c$
are well described by the GL theory, where $T_c$ is the critical
temperature when the applied magnetic field is zero.
The free energy is represented
by the order parameter ${\cal O}$ and a vector potential $\bmA$ as
\begin{align} 
& F=
\int d\bmx \left[c(T)|D{\cal O}|^2-a(T)|{\cal O}|^2
+\frac{b(T)}{2}|{\cal O}|^4\right], 
 \nonumber \\
& D:=\nabla-ie_\ast \bmA, 
\end{align}
where $e_\ast$ is the effective charge coupled to
the vector potential $\bmA$.
Here, we assume that the parameter $a(T)$ changes
from negative to positive at $T=T_c$ as $T$ decreases, 
while the other parameters, $b$ and $c$ remain positive near $T=T_c$.
The current $\bm j$ is given by 
\begin{align}
\label{TDGL-current}
   \bmj
  &= - \frac{\delta F}{\delta \bmA}
  = 2 e_\ast\, c(T) \left( \Im[{\cal O}^\dagger \nabla {\cal O}]
  - e_\ast\, |{\cal O}|^2 \bmA \right)~.
\end{align}

The TDGL equation describing the non-equilibrium states
is given by a kinetic coefficient $\Gamma(T)$ as
\begin{align}
\label{TDGL-eq}
  & ( \p_t - i e_\ast \phi){\cal O}
  = -
  \Gamma(T)\, \frac{\delta F}{\delta {\cal O}^\dagger}
\nonumber \\
  &\hspace*{1.0truecm}
  = \Gamma(T)\left(c(T)D^2+a(T)-b(T)|{\cal O}|^2\right){\cal O}~,
\end{align}
where $\phi$ is the electric potential.

We first consider the superconducting state in the absence of electric 
and magnetic fields. 
Setting the l.~h.~s. of (\ref{TDGL-eq}) to zero, we obtain
a stationary homogeneous solution for $T<T_c$ as
\begin{align}
\label{App:homogeneous-sol}
  & {\cal O} = {\cal O}_0(T) := \sqrt{\frac{a(T)}{b(T)}}
  \sim \left(1-\frac{T}{T_c}\right)^{1/2} 
  =: \epsilon_T^{1/2} 
\end{align}
near the critical temperature.

From the TDGL equation (\ref{TDGL-eq}),
the perturbation $\delta \calO := {\cal O} - {\cal O}_0$
around the homogeneous condensate (\ref{App:homogeneous-sol})
satisfies the dispersion relation
\begin{align}
\label{App:QNM}
  & \omega
  = - i\, c(T)\, \Gamma(T)\,
    \left( k^2 + \frac{2\, a(T)}{c(T)} \right)~,
\end{align}
where we set
$\delta \calO \propto e^{- i \omega t + i \bmk \cdot \bmx}$.
This yields the correlation length $\xi$ from the wave number $k_*$
generating the static perturbation, i.e., $\omega=0$ as
\begin{align}
\label{App:correlation length}
  & \xi^2 = - \frac{1}{k_*^2} = \frac{c(T)}{2a(T)}~.
\end{align}

Next we consider the magnetic and electric response.
Applying the infinitesimal magnetic field on the homogeneous condensate
(\ref{App:homogeneous-sol}),
the London current is generated by the vector potential $\bm A$ as
\cite{parks} 
\begin{align}
\label{App-London}
  & \bmj =
  - 2\, e_\ast^2\, c(T)\, |{\cal O}_0|^2\, \bmA~.
\end{align}
When the field strength increases beyond a critical value $B_{c1}$,
the external magnetic field begins to penetrate into the superconductor
and vortices appear.
At $B=B_{c2}~(>B_{c1})$, the second order phase transition occurs
and the superconductivity disappears.
Just below the upper critical value $B_{c2}$,
a triangular lattice appears
since it is thermodynamically most favorable solution
(in details, see Ref.~\cite{parks}).

For simplicity, we consider the following gauge fields
generating the upper critical magnetic field $B_{c2}$
and a small electric field $E$ in the $x$-direction:
\begin{align}
\label{eq:gauge_in_TDGL}
\bmA=-(B_{c2}\,y+E\,t)dx, \qquad \phi=0.
\end{align}
Substituting an ansatz
\begin{align}
  & {\cal O}(\bm x, t)
  = \int dp~C(p) e^{ipx} \xi(\hy)~,
& & \hat{y} := y + \frac{E}{B_{c2}}\, t
\end{align}
into Eq.~(\ref{TDGL-eq}), we obtain
\begin{align}
\label{App:eq-xi}
   \frac{E}{\Gamma(T) B_{c2}}\, \odiff{\xi}{\hy}
  &= c(T) \left[ \odiffII{}{\hy}
  - ( p + e_\ast B_{c2}\, \hy )^2 \right] \xi
\nonumber \\
  &+ a(T)\, \xi
\end{align}
just below the critical temperature $T_c$.
Here, we neglected the quadratic term with respect to $\xi$
because $T$ is very close to $T_c$~($\epsilon_T$ is very small).
Introducing a new variable $\chi$ as
\begin{align}
  & \xi(\hat{y})
  = \exp\left( \frac{E}{2 c(T) \Gamma(T) B_{c2}}\, \hy \right)
  \chi(\hy)~,
\end{align} 
Eq.~(\ref{App:eq-xi}) can be simplified as
\begin{align}
\label{App:eq-chi}
  & \left( \frac{d^2}{dY^2} - \frac{Y^2}{4} \right) \chi
  = - \frac{a(T)}{2e_\ast c(T) B_{c2}}\, \chi + O(E^2)~,
\\
  & Y
  := \sqrt{2e_\ast B_{c2}}
  \left(\hy + \frac{p}{e_\ast B_{c2}} \right)~.
\end{align}
Neglecting the square term for small $E$ and  
comparing Eq.~(\ref{App:eq-chi}) with Eq.~(\ref{eq:D_n}), 
we finally obtain a vortex flow solution with the lowest energy as 
\begin{align} 
\label{App:sol-m}
   \calO
  &= \int dp~C(p)\, \exp\left( i p x
    + \frac{E}{2 c(T) \Gamma(T) B_{c2}}\, \hy \right)
\nonumber \\
  &\times \exp\left[ - \frac{e_\ast B_{c2}}{2}
    \left( \hy + \frac{p}{e_\ast B_{c2}} \right)^2
    \right]
\end{align}
with the relation
\begin{align}
\label{App:relation}
  & a(T) =e_* c(T) B_{c2}~.
\end{align}

Substituting Eq.~(\ref{App:sol-m}) into Eq.~(\ref{TDGL-current}), 
we obtain
\begin{subequations}
\label{eq:vortex_flow-current}
\begin{align}
  & j_x = - c(T)\, e_\ast\, \partial_y | \calO |^2
  + \frac{e_\ast\,E}{\Gamma(T)\,B_{c2}}\, | \calO |^2~,
\label{eq:j_x} \\
  & j_y = c(T)\, e_\ast\, \partial_x | \calO |^2~.
\label{eq:j_y}
\end{align}
\end{subequations}
%
%
%
Averaging the values of the current over the ${\bm x}$-space, 
we find that only $\overline{j_x}$ is non-zero:
\begin{align}
\label{App:current}
  & \overline{j_x}
  = \frac{ e_\ast E}{\Gamma(T)B_{c2}}\,
  \overline{|{\cal O}|^2}~.
\end{align} 
Eq.~(\ref{App:current}) shows the Ohmic dissipation,
as the electric field is applied in the $x$-direction, i.e.,
${\bm E}=-\p_t{\bm A}=Edx$.
\section{The derivation of the coefficients in the TDGL equation}
In the following, we will determine the parameters $a$, $b$, and $c$ in Eq.~(\ref{TDGL-eq}) 
just below the critical temperature $T_c$ from the correlation 
length and the London equation calculated in the holographic superconductor model~\cite{Maeda:2008ir}.  

In the absence of magnetic field, the scalar field $\Psi$ and the gauge potential $A_t$ 
can be expanded as   
\begin{align}
& \Psi(u)=\frac{\epsilon_T^{1/2}}{L}(\tilde{\Psi}_1(u)+\epsilon_T\tilde{\Psi}_2(u)+\cdots), 
\nonumber \\
& A_t(u)=\alpha(T)\{q_c(1-u)+\epsilon_T\tilde{\Phi}_1(u)+O(\epsilon_T^2)\}, 
\end{align}
where $q_c=\mu/\alpha(T_c)$. Then, as shown in Ref.~\cite{Maeda:2008ir}, 
the equations of motion for $\tilde{\Psi}_1(u)$ and $\tilde{\Phi}_1(u)$ are written by 
\begin{align}
\label{App:psi-phi}
{\cal L}_\psi\,\tilde{\Psi}_1=0, \qquad 
\frac{d^2\tilde{\Phi}_1(u)}{du^2}=s(u):=\frac{2q_c|\tilde{\Psi}_1|^2}{u^2(1+u+u^2)}, 
\end{align}
where ${\cal L}_\psi$ is defined by ${\cal L}$ in Eq.~(\ref{eq:def-calL}) as 
${\cal L}_\psi:=-{\cal L}|_{T=T_c}$. 
As mentioned in Sec.~II, we consider the boundary conditions for $\tilde{\Psi}_1$:
\begin{align}
\label{App-bc-psi}
\lim_{u\to 0}\tilde{\Psi}_1(u)=O(u^2), \qquad \tilde{\Psi}_1(u=1)=\mbox{regular}. 
\end{align} 
We derive the boundary conditions for $\tilde{\Phi}_1(u)$ from 
the requirement that the chemical potential $\mu=A_t(0)$ is 
fixed under the variation of the temperature:
\begin{align}
\label{App:phi1-bc}
\lim_{u\to 0}\tilde{\Phi}_1(u)=q_c, \qquad \tilde{\Phi}_1(u=1)=0. 
\end{align}
Here, the latter condition is the regularity condition at the horizon. 
The formal solution $\tilde{\Phi}_1$ satisfying the boundary conditions 
is given by 
\begin{align}
\label{app-phi1}
& \tilde{\Phi}_1(u)=-u\int^1_us(v)(1-v)dv \nonumber \\
&-(1-u)\int^u_0s(v)vdv+q_c(1-u). 
\end{align} 

In terms of $\tilde{\Psi}_1$ and $\tilde{\Phi}_1$, the correlation length $\xi$ is 
represented by 
\begin{align}
\label{correlation-AdS}
\xi^2\simeq\frac{\epsilon_T^{-1}}{\alpha(T_c)}\frac{D}{N} 
\end{align} 
in the limit $T\to T_c$~\footnote{Under the asymptotic boundary condition in 
Eq.~(\ref{App:phi1-bc}), $\epsilon:=q/q_c-1$ defined in Ref.~\cite{Maeda:2008ir} 
is equal to $\epsilon_T$.}, where 
\begin{align}
& N=2\int^1_0du\left(\frac{d\tilde{\Phi}_1}{du}+\tilde{\Phi}_1(0)\right)^2, \nonumber \\
& D=\int^1_0du \frac{\tilde{\Psi}_1^2(u)}{u^2}. 
\end{align}
Since $\tilde{\Psi}_1$ is a solution of the linear equation~(\ref{App:psi-phi}), its  
amplitude is obtained from the next order equation:
\begin{align}
{\cal L}_\psi\tilde{\Psi}_2=-\frac{2q_c\tilde{\Phi}_1\tilde{\Psi}_1}{1+u+u^2}. 
\end{align}  
The boundary conditions for $\tilde{\Psi}_2$ are the same as the ones for 
$\tilde{\Psi}_1$~(\ref{App-bc-psi}). Under the boundary conditions, 
${\cal L}$ is Hermitian for the inner product~(\ref{eq:def-IP-AdS_rad}). 
This yields 
\begin{align}
\label{app-cond-psi2}
& 0=\langle {\cal L}_\psi\tilde{\Psi}_2,\,\tilde{\Psi}_1 \rangle=
\langle \tilde{\Psi}_2,\,{\cal L}_\psi\tilde{\Psi}_1 \rangle \nonumber \\
&=-2q_c\left\langle \tilde{\Psi}_1,\,\frac{\tilde{\Phi}_1\tilde{\Psi}_1}{1+u+u^2}\right\rangle. 
\end{align}
Substituting Eq.~(\ref{app-phi1}) into Eq.~(\ref{app-cond-psi2}), 
we obtain the amplitude $A$ defined by $A:=\tilde{\Psi}_1/\tilde{\psi}_1$ for a normalized 
solution $\tilde{\psi}_1$ satisfying $\tilde{\psi}_1(1)=1$ as   
\begin{align}
& A^2=\frac{\Sigma_2(0)}
{2\left[\int^1_0\Sigma_1^2(u)du-\left(\int^1_0\Sigma_1(u)du\right)^2\right]}, \nonumber \\
& \Sigma_n(u)=\int^1_u\frac{(1-v)^n\tilde{\psi}_1^2(v)}{v^2h(v)}dv \qquad (n=1,\,2). 
\end{align}  

Numerical calculation determines the value of the amplitude $A$ and hence 
$c_2$ in Eq.~(\ref{eq:bc-bdy}) is evaluated as
\begin{align}
\label{App:coe-c2}
c_2\simeq\lim_{u\to 0}\frac{\tilde{\Psi}_1}{u^2}\simeq \frac{6.55\,\epsilon_T^{1/2}}{L}. 
\end{align}
$\xi$ in Eq.~(\ref{correlation-AdS}) is also evaluated as 
\begin{align}
\label{App:correlation length1}
\xi\simeq 0.0635\times\frac{\epsilon_T^{-1/2}}{T_c}.  
\end{align}

For the vector potential ${\bm A}$ generating small magnetic field, the London equation 
just below $T=T_c$~\cite{Maeda:2008ir} is evaluated as 
\begin{align}
\label{App-London1}
  & \lim_{T\to T_c}\Exp{{\bm J}}
  \simeq - \frac{L^4\, \alpha(T_c)}{2\, \kappa_4^2\, e^2}\,
  \left( 2 \int^1_0 du~\frac{| \Psi |^2}{c_2^2\,u^2} \right) c_2^2\,{\bm A} 
\nonumber \\ 
  &  \simeq - \frac{L^2\, \alpha(T_c)}{2\, \kappa_4^2\, e^2}\,
  \left( 2 \int^1_0 du~\frac{\epsilon_T| \tilde{\Psi}_1 |^2}{c_2^2\,u^2} \right) c_2^2\,{\bm A} 
  \nonumber \\
&\simeq - \frac{2\, \kappa_4^2}{L^2 T_c^3} 
  \times 1.172\times 10^{-3}\,|\Exp{{\cal O}_2}|^2 {\bm A}, 
\end{align}
where we used Eqs.~(\ref{eq:def-h_alpha}) and~(\ref{eq:expectation-o}) to 
derive the third equality. Comparing Eq.~(\ref{App-London1}) 
with Eq.~(\ref{App-London}) and identifying $\Exp{{\cal O}_2}$ with ${\cal O}_0$, 
we obtain the coefficient $c$ in the TDGL equation~(\ref{TDGL-eq}) just below 
$T_c$ as
\begin{align}
\label{App:coe-c}
c(T_c)\simeq \frac{2\, \kappa_4^2}{L^2}
\times \frac{5.85\times 10^{-4}}{T_c^3}, 
\end{align}
where we used the fact, $e_*=1$ derived in Sec.~IV. 
Substitution of Eqs.~(\ref{App:correlation length1}) and (\ref{App:coe-c}) into  
Eq.~(\ref{App:correlation length}) yields the coefficient $a$ just below $T_c$ as
\begin{align}
\label{App:coe-a}
a(T\to T_c)\sim \frac{2\kappa_4^2}{L^2}\times \frac{0.0726\,\epsilon_T}{T_c}.  
\end{align}
$b$ is also evaluated 
from Eqs.~(\ref{eq:expectation-o}), (\ref{App:homogeneous-sol}), and (\ref{App:coe-c2}) as
\begin{align}
b(T_c)=\left(\frac{2\kappa_4^2}{L^2}\right)^3e^{2}\times \frac{2.75\times 10^{-6}}{T_c^5}. 
\end{align}

\section{Ohmic dissipation}
Since the bulk spacetime possesses Killing vector
$\xi^\mu = (\partial_t)^\mu = (\partial_v)^\mu$,
the energy-momentum tensor of the form
\begin{align}
   T_{\mu\nu}
  &= \frac{L^2}{2 \kappa_4^2 e^2}\,
  \Bigg[~F_{\mu\lambda} F_\nu{ }^\lambda
  + 2\, \Re\left[\, \big( D_{\mu} \Psi \big)^\dagger
    \big( D_{\nu} \Psi \big)\, \right]
\nonumber \\
  &- g_{\mu\nu} \left( \frac{F^2}{4} + \big\vert\, D\Psi\, \big\vert^2
    + m^2\, | \Psi |^2 \right)~\Bigg]~,
\label{eq:EM_tensor}
\end{align}
satisfies the conservation law,
$\nabla_\mu \left( T^\mu{}_\nu\, \xi^\nu \right)=0$.
Thus, we obtain 
\begin{align}
   0
  &= \int d^{4}x~\sqrt{-g}\,
    \nabla_\mu \left( T^\mu{}_\nu\, \xi^\nu \right)
\nonumber \\
  &= \int_{\Sigma_f} d\Sigma_\mu\, T^\mu{}_\nu\, \xi^\nu
  + \int_{\Sigma_i} d\Sigma_\mu\, T^\mu{}_\nu\, \xi^\nu
\nonumber \\
  &+ \int_{\calH} d\Sigma_\mu\, T^\mu{}_\nu\, \xi^\nu
  + \int_{\text{bdy}} d\Sigma_\mu\, T^\mu{}_\nu\, \xi^\nu~,
\label{eq:conserve_eqn}
\end{align}
where, $\Sigma_{f,i}$, $\calH$, and bdy represent
$v = \text{const.}$ null hypersurfaces, the null hypersurface at 
the black hole horizon, and the timelike AdS boundary, respectively.
Since the spatially averaged value of $T_{\mu\nu}$ does not depend
on $v$, the first and the second terms in the second line 
of Eq.~(\ref{eq:conserve_eqn}) cancel each other.
This implies
\begin{align}
   - \int_{\calH} d\Sigma_a\, T^a{}_b\, \xi^b
  &= \int_{\text{bdy}} d\Sigma_a\, T^a{}_b\, \xi^b~.
\label{eq:conserve_eqn-stationary}
\end{align}

Due to the rapid fall-off condition for $\Psi$ (\ref{boundary-con}),
only the $U(1)$ bulk gauge field contributes to the boundary term
in the above equation as the Ohmic dissipation:
\begin{align*}
   \int_{\text{bdy}} d\Sigma_a\, T^a{}_b\, \xi^b
  &= \int_{\text{bdy}} dv\, d^2x \Exp{J^i}\, E_i~.
\end{align*}
Here, we used the fact
\begin{align*}
  & \Exp{J^i}
  = \frac{L^2}{2 \kappa_4^2 e^2}\, \sqrt{- g}~F^{u i}
  \, \Big\vert_{u=0},
\\
  & E_i
  = - F_{t i}\, \big\vert_{u=0}
  = - F_{v i}\, \big\vert_{u=0}
& & (i = x, y)~.
\end{align*}
%
%
%
Hence, Eq.~(\ref{eq:conserve_eqn-stationary}) is reduced to
\begin{align}
  & \frac{L^3\, \alpha}{2 \kappa_4^2 e^2}
  \int_{\calH} dv\, d^2x~\left( \delta^{ij}\, F_{vi} F_{vj}
    + 2\, (L\, \alpha)^{2}\, \big\vert\, D_v \Psi\, \big\vert^2 \right)
%
\nonumber \\
  &= \int_{\text{bdy}} 
  dv\, d^2x~\Exp{J^i}\, E_i~.
\label{eq:conserve_eqn-stationary-concrete}
\end{align}

We can extract the subleading terms at $O(\epsilon)$ from above. 
Noting
%
\begin{align*}
   \int dx dy~f_{vj}
  &= \int dx dy~2\, \partial_{[v} a_{j]}
  = - \int dx dy~\partial_j a_v
  = 0~,
\end{align*}
we obtain
\begin{align}
  & \frac{L^5\, \alpha^3}{\kappa_4^2 e^2}
  \int_{\calH} dv\, d^2x~\big\vert\, \bmD_v \Psi \, \big\vert^2
  = \int_{\text{bdy}} 
  dv\, d^2x~\delta\Exp{J^i}\, E_i~.
\label{eq:conserve_eqn-stationary-concrete-perturbed}
\end{align}
%



\begin{thebibliography}{99}

\bibitem{Maldacena:1997re}
  J.~M.~Maldacena,
  ``The large N limit of superconformal field theories and supergravity,"
  Adv.\ Theor.\ Math.\ Phys.\  {\bf 2} (1998) 231
  [Int.\ J.\ Theor.\ Phys.\  {\bf 38} (1999) 1113]
  [arXiv:hep-th/9711200].  

\bibitem{Gubser:2008px}
  S.~S.~Gubser,
  ``Breaking an Abelian gauge symmetry near a black hole horizon,''
  Phys.\ Rev.\  D {\bf 78} (2008) 065034
  [arXiv:0801.2977 [hep-th]].

\bibitem{Hartnoll:2008vx}
  S.~A.~Hartnoll, C.~P.~Herzog and G.~T.~Horowitz,
  ``Building a Holographic Superconductor,''
  Phys.\ Rev.\ Lett.\  {\bf 101}, 031601 (2008)
  [arXiv:0803.3295 [hep-th]].

\bibitem{Albash:2009ix}
  T.~Albash and C.~V.~Johnson,
  ``Phases of Holographic Superconductors in an External Magnetic Field,''
  arXiv:0906.0519 [hep-th].

\bibitem{Albash:2009iq}
  T.~Albash and C.~V.~Johnson,
  ``Vortex and Droplet Engineering in Holographic Superconductors,''
  Phys.\ Rev.\  D {\bf 80} (2009) 126009
  [arXiv:0906.1795 [hep-th]].
  
\bibitem{Montull:2009fe}
  M.~Montull, A.~Pomarol and P.~J.~Silva,
  ``The Holographic Superconductor Vortex,''
  Phys.\ Rev.\ Lett.\  {\bf 103} (2009) 091601
  [arXiv:0906.2396 [hep-th]].

\bibitem{Maeda:2009vf}
  K.~Maeda, M.~Natsuume and T.~Okamura,
  ``Vortex lattice for a holographic superconductor,''
  Phys.\ Rev.\  D {\bf 81}, 026002 (2010)
  [arXiv:0910.4475 [hep-th]].



\bibitem{Maeda:2008ir}
  K.~Maeda and T.~Okamura,
  ``Characteristic length of an AdS/CFT superconductor,''
  Phys.\ Rev.\  D {\bf 78}, 106006 (2008)
  [arXiv:0809.3079 [hep-th]].


\bibitem{Hartnoll:2008kx}
  S.~A.~Hartnoll, C.~P.~Herzog and G.~T.~Horowitz,
  ``Holographic Superconductors,''
  JHEP {\bf 0812}, 015 (2008)
  [arXiv:0810.1563 [hep-th]].

\bibitem{Maeda:2009wv}
  K.~Maeda, M.~Natsuume and T.~Okamura,
  ``Universality class of holographic superconductors,''
  Phys.\ Rev.\  D {\bf 79}, 126004 (2009)
  [arXiv:0904.1914 [hep-th]].

\bibitem{AmadoKaminskiLandsteiner2009}
``Hydrodynamics of Holographic Superconductors," 
Irene Amado, Matthias Kaminski, and Karl Landsteiner, 
JHEP {\bf 0905}, 021 (2009) 
[arXiv:0903.2209 [hep-th]]. 

\bibitem{mkf2009}
K.~Maeda, S.~Fujii, and J.~Koga, 
``The final fate of instability of Reissner-Nordstrom-anti-de Sitter 
black holes by charged complex scalar fields,''
Phys.\ Rev.\ D {\bf 81}, 124020 (2010) 
[arXiv:1003.2689 [gr-qc]].   

\bibitem{Herzog:2008he}
  C.~P.~Herzog, P.~K.~Kovtun and D.~T.~Son,
  ``Holographic model of superfluidity,''
  Phys.\ Rev.\  D {\bf 79}, 066002 (2009)
  [arXiv:0809.4870 [hep-th]].

\bibitem{Albash:2008iv}
T.~Albash and C.~V.~Johnson, 
``A Holographic Superconductor in an External Magnetic Field,'' 
JHEP {\bf 0809}, 121 (2008). 

\bibitem{parks}
R.~D.~Parks, {\it Superconductivity} (Marcel Dekker Inc., New
York, 1969);
A.~A.~Abrikosov, {\it Fundamentals of the Theory of Metals}
(North-Holland, New York, 1988);
M. Tinkham, {\it Introduction to Superconductivity} (McGraw-Hill Inc., New York, 1996).

\bibitem{Maeda:2010br}
  K.~Maeda, M.~Natsuume and T.~Okamura,
  ``On two pieces of folklore in the AdS/CFT duality,''
  Phys.\ Rev.\  D {\bf 82}, 046002 (2010)
  [arXiv:1005.2431 [hep-th]].





\end{thebibliography}
\end{document}